\documentclass[fleqn,usenatbib]{mnras}

\usepackage{mathptmx}
\usepackage{txfonts}

\usepackage[T1]{fontenc}

\usepackage{graphicx}	
\usepackage{color}
\usepackage{float}
\usepackage{ulem}

\def\apj {ApJ}
\def\apjl {ApJL}
\def\apjs {ApJS}
\def\aj {AJ}
\def\aap {A\&A}
\def\mnras {MNRAS}
\def\araa {ARA\&A}
\def\nat {Nature}
\def\hmsun{h^{-1} {\rm{M_\odot}}}
\def\sag {\textsc {sag}}

\def\mdpl {\textsc {mdpl}\oldstylenums{2}}

\title[Backsplash galaxies]{
Backsplash galaxies and their impact on galaxy evolution: a three-stage, four-type perspective}
\author[Ruiz et al.]{\parbox[t]{\textwidth}{
Andr\'es N. Ruiz,$^{1,2}$\thanks{andres.ruiz@unc.edu.ar} 
H\'ector J. Mart\'inez,$^{1,2}$
Valeria Coenda,$^{1,2}$ 
Hern\'an Muriel,$^{1,2}$
Sof\'ia A. Cora,$^{3,4}$
Mart\'in de los Rios,$^{5,6}$
and Cristian A. Vega-Mart\'inez$^{7,8}$
}
\\
\\ 
$^{1}$Instituto de Astronom\'ia Te\'orica y Experimental (CONICET-UNC), 
Laprida 854, X5000BGR, C\'ordoba, Argentina\\
$^{2}$Observatorio Astron\'omico, Universidad Nacional de C\'ordoba,
Laprida 854, X5000BGR, C\'ordoba, Argentina\\
$^3$ Instituto de Astrof\'isica de La Plata (CONICET-UNLP), Observatorio 
Astron\'omico, Paseo del Bosque S/N, B1900FWA, La Plata, Argentina\\
$^4$ Facultad de Ciencias Astron\'omicas y Geof\'isicas, Universidad 
Nacional de La Plata, Observatorio Astron\'omico, Paseo del Bosque S/N, 
B1900FWA, La Plata, Argentina\\
$^{5}$ Departamento de F\'isica Te\'orica, Universidad Aut\'onoma de 
Madrid, 28049 Madrid, Spain\\
$^{6}$ Instituto de F\'isica Te\'orica, IFT-UAM/CSIC, C/ Nicolás 
Cabrera 13-15, Universidad Autónoma de Madrid, Cantoblanco, Madrid 
28049, Spain\\
$^7$ Instituto Multidisciplinario de Investigaci\'on y
Postgrado, Universidad de La Serena, Ra\'ul Bitr\'an 1305, La 
Serena, Chile\\
$^8$ Departamento de Astronom\'ia, Universidad de La Serena, 
Av. Juan Cisternas 1200 Norte, La Serena, Chile
}

\begin{document}
\label{firstpage}
\pagerange{\pageref{firstpage}--\pageref{lastpage}}
\maketitle

\begin{abstract}
We study the population of backsplash galaxies at $z=0$ in the outskirts of
massive, isolated clusters of galaxies taken from the \mdpl-\sag~ 
semi-analytic catalogue.  We consider four types of backsplash galaxies
according to whether they are forming stars or passive at three stages
in their lifetimes: before entering the cluster, during their first
incursion through the cluster, and after they exit the cluster. We
analyse several geometric, dynamic, and astrophysical aspects of the
four types at the three stages.  Galaxies that form stars at all stages
account for the majority of the backsplash population ($58\%$) 
and have stellar masses typically below $M_\star\sim 3\times 10^{10} \hmsun$ 
that avoid the innermost cluster's regions and are only mildly affected 
by it. In a similar mass range, galaxies that become passive after
exiting the cluster ($26\%$) follow orbits characterised by small
pericentric distance and a strong deflection by the cluster potential
well while suffering a strong loss of both dark matter and gas
content. Only a small fraction of our sample ($4\%$) become
passive while orbiting inside the cluster. These galaxies have
experienced heavy pre-processing and the cluster's tidal stripping and
ram pressure provide the final blow to their star formation.  Finally,
galaxies that are passive before entering the cluster for the first
time ($12\%$) are typically massive and are not affected
significantly by the cluster. Using the bulge/total mass ratio as a
proxy for morphology, we find that a single incursion through a cluster
do not result in significant morphological changes in all four types.
\end{abstract}



\section{Introduction}
\label{sec:intro}

For several decades, the study of galaxy clusters and their impact on the
galaxy evolution has remained as a  captivating research subject. The process
of galaxy evolution within clusters involves a myriad of physical mechanisms
that operate across varying spatial and temporal scales. Both internal and
environmental factors exert significant influence on the characteristics of
galaxies, giving rise to a wide range of possible transformations and
adaptations. Supernovae (e.g. \citealt{Stringer:2012}, \citealt{Bower:2012},
\citealt{Christensen:2016}), active nuclei (e.g.  \citealt{Nandra:2007},
\citealt{Hasinger:2008}, \citealt{Silverman:2008}, \citealt{Cimatti:2013}), and
stellar feedback (e.g.  \citealt{DallaV:2008}, \citealt{Hopkins:2012}) are
among the internal mechanisms that impact galaxies. On the other hand,
properties of galaxies, such as star formation (e.g.
\citealt{Hashimoto,Mateus:2004,BlantonMoustakas:2009,Welikala:2008,Schaefer:2017,Coenda:2019}),
morphology (e.g. \citealt*{dressler80,Whitmore:1993,Dominguez:2001};
\citealt{Weinmann:2006,Bamford:2009, Paulino-Afonso:2019}), luminosity (e.g.
\citealt*{Adami:1998}; \citealt{Coenda:2006}), color (e.g. \citealt{Blanton05,
martinez06}; \citealt*{martinez08}), age (e.g.  \citealt{Thomas:2005},
\citealt{Cooper:2010a}, \citealt{Zheng:2017}), are significantly influenced by
their environment. 

Several mechanisms impact galaxies in clusters, some of which cause gas
depletion and the halt of star formation. Galaxies that move at high velocities
through the intra-cluster medium (ICM) experience ram pressure stripping (e.g.
\citealt{GG:1972}; \citealt*{Abadi:1999,Book:2010, Steinhauser:2016}), which
removes a significant portion of their cold gas and reduces their star
formation rate (SFR). Moreover, a galaxy's trip through the ICM can result in
the removal of its warm gas, a phenomenon known as starvation
(\citealt{Larson:1980,McCarthy:2008,Bekki:2009,Bahe:2013,Vijayaraghavan:2015}).
Starvation inhibits future star formation by cutting off the supply of gas that
cools from the galaxy's halo. Tidal stripping from the cluster potential is
another mechanism that eliminates the gas supply
(\citealt*{Zwicky:1951,Gnedin:2003a,Villalobos:2014}). In contrast, in
intermediate-density areas such as cluster outskirts and groups, harassment is
a more effective galaxy-galaxy interaction mechanism (e.g.
\citealt{Moore:1996}; \citealt{Moore:1998}, \citealt{Gnedin:2003b}). It causes
both gas depletion and morphological transformations. Tidal stripping from
galaxy-galaxy encounters can truncate galaxies, mainly disks, and result in
spheroid-dominated galaxies (e.g. \citealt{Smith:2015}). Morphological
evolution primarily depends on mergers (e.g. \citealt{Toomre:1977, Barnes:1992,
DiMatteo:2007, Martin:2018}), with gas-rich minor mergers producing massive
disks \citep{Jackson:2022}, and with major mergers resulting in spheroidal
systems \citep{Navarro:1994}.

Galaxies may encounter various environmental conditions and be subjected to one
or more of the mechanisms discussed earlier at different stages of their life
cycle. As galaxies fall towards clusters, they can experience different
physical processes depending on whether they are part of a group (e.g.
\citealt{McGee:2009,deLucia:2012,Wetzel:2013}; \citealt*{Hou:2014}), falling
from the field (e.g. \citealt{Berrier:2009}), or through filament streams (e.g.
\citealt{Colberg:1999}; \citealt{Ebeling:2004,Martinez16}; \citealt{Rost:2020,
Kuchner:2022}). Before entering the cluster, galaxies can undergo several
physical transformations due to these processes, which are collectively called
pre-processing (e.g. \citealt{Mihos:2004,Fujita:2004}).

After being incorporated into a cluster, galaxies have two possible outcomes:
they can either remain bound to the cluster's gravitational field or their
trajectory can carry them away from the cluster, spanning distances of up to
several $R_{200}$, the radius enclosing 200 times the mean density of the
Universe (e.g. \citealt{Mamon:2004};
\citealt{Gill:2005,Rines:2006,Aguerri:2010,Muriel:2014, Casey:2023}).
Eventually, these galaxies will turn around and fall back into the cluster
during a subsequent infall. This unique population of galaxies is known as
backsplash galaxies \citep{Balogh:2000}. These particular galaxies can be used
as laboratories to explore the impact of the cluster environment in galaxy
properties, and to disentangle which stage is more important in their
lifetimes.  However, after these galaxies left the cluster, they may show
characteristics of "post-processing" that are difficult to distinguish from
"pre-processing" signatures. In addition, observational studies have
demonstrated that environmental effects may extend beyond the halo boundary to
impact both baryonic components (e.g. \citealt{Wetzel:2012, Lu:2012}) and dark
matter haloes (e.g.  \citealt{Behroozi:2014}).

In the last years there have been many works about backsplash galaxies from a
theoretical perspective. \citet{Benavides:2021}, using the
\textsc{IllustrisTNG} hydrodynamical simulations
\citep{illustrisTNG_1,illustrisTNG_2,illustrisTNG_3,illustrisTNG_4,illustrisTNG_5},
have found that backsplash galaxies that were in the past satellite of another
group or cluster, can be the origin of quenched ultra-diffuse galaxies.
\citet{Kuchner:2022} explores the concept of backsplash galaxies that are
falling into clusters along filaments. For the authors, these are galaxies
outside the $R_{200}$ that remain gravitationally bound to the cluster, and
they may have made several orbits around the potential center. The study
reveals that between 30 to 60 per cent of filament galaxies are classified as
backsplash galaxies. Interestingly, backsplash galaxies return to the cluster
after deviating significantly from their initial trajectory upon entry,
particularly in more relaxed clusters. They do not exhibit a preferred location
with respect to filaments and are unable to collapse and form filaments
themselves.  Several studies of backsplash galaxies around clusters have been
carried out using \textsc{TheThreeHundred} project \citep{Cui:2018}, a suite of
hydrodynamical resimulations of galaxy clusters.  \citet{haggar_2020} found
that the fraction of backsplash galaxies inhabiting a shell between $R_{200}$
and $2R_{200}$ vary from 21 to 85 per cent, with a mean value of 58 per cent,
and that this fraction is dependent of the dynamical state of the cluster.
\citet{knebe_2020} perform a detailed study of the shape and alignment of
galaxies around clusters, finding that backsplash galaxies have a larger radial
alignment and more spherical shapes than the infalling population of galaxies.
\citet{Hough:2023} discovered that approximately 65 per cent of quenched
galaxies situated near clusters are backsplash galaxies. This suggests that a
combination of ram pressure stripping during the pre-processing stage and
within the cluster is required to suppress star formation.  Recently,
\citet{Borrow:2023}, use \textsc{IllustrisTNG} simulations to study backsplash
galaxies around 1302 isolated galaxy clusters with mass $10^{13.0}<M_{200}/{\rm
M}_{\odot}<10^{15.5}$. Their studies show that backsplash galaxies exhibit
distinct characteristics compared to field galaxies, such as low gas fractions,
high mass-to-light ratios, large stellar sizes, and low black hole occupation
fractions. 

The aim of this study is to investigate the existence and consequences of
pre-processing in backsplash galaxies that have yet to cross the virial radius,
as well as the effects of clusters on their subsequent evolution, once they are
situated in the outskirts. Specifically, this research delves into the life
cycle of these galaxies starting from 2 Gyr prior to their initial crossing of
the virial radius.  The present paper is organised as follows. In Sec.
\ref{sec:data} we describe the simulated galaxy catalogue and define the types
and stages considered. Dynamical and astrophysical properties of BS galaxies
are analysed in Sec. \ref{sec:dynamics} and \ref{sec:astrophysics},
respectively. Finally, in Sec. \ref{sec:conclusions} we present the main
remarks of our work. 


\section{Data}
\label{sec:data}

To construct our sample of simulated galaxy clusters, we have combined dark
matter-only simulations of regions that contain cluster-like haloes at $z = 0$
taken from the \mdpl~cosmological simulation, along with the semi-analytic
model of galaxy formation \sag. In this regard, we will first provide a brief
overview of the dark matter simulation and the semi-analytic model, followed by
a description of the selection criteria employed to construct the galaxy sample
in and around clusters.


\subsection{The \mdpl~cosmological simulation}
\label{sec:simu}

The \mdpl~cosmological simulation is one of several simulations within the
\textsc{MultiDark} suite \citep{riebe_multidark_2013,klypin_mdpl2_2016}. This
simulation contains $3840^3$ dark matter particles in a comoving cubic box
measuring $1 h^{-1} {\rm Gpc}$ in length on each side. Each particle has a mass
of $m_{\rm p} = 1.51 \times 10^9 \hmsun$. The simulation assumes a flat
$\Lambda$CDM cosmology with $\Omega_\mathrm{m} = 0.307$, $h = 0.678$, $n =
0.96$, and $\sigma_8 = 0.823$, which is consistent with measurements made by
\citet{planck_cosmology_2014,planck_cosmology_2016}. Using {\sc
gadget-}\oldstylenums{2} \citep{springel_gadget2}, the simulation tracks the
dynamical evolution of dark matter particles from an initial redshift $z =
120$. Over the course of the simulation, $126$ snapshots were recorded between
redshifts $z = 17$ and $z = 0$. 

Dark matter haloes and subhaloes were identified using the {\sc Rockstar}
phase-space halo finder \citep{behroozi_rockstar_2013}, and merger trees were
constructed using {\sc ConsistentTrees} \citep{behroozi_trees_2013}. The
halo/subhalo catalogues and merger trees used in this study are publicly
accessible through the {\sc
CosmoSim}\footnote{https://www.cosmosim.org/cms/simulations/mdpl2/} and {\sc
Skies$\& $Universes
}\footnote{http://skiesanduniverses.org/Simulations/MultiDark/} databases, and
form the foundation of the semi-analytic model for generating the galaxy
population.

\subsection{The \sag~model}
\label{sec:sag}

The semi-analytic model of galaxy formation and evolution \sag~(acronym for
Semi-Analytic Galaxies) has its roots in the model presented by
\citet{Springel+2001}; the most recent version is described in
\citet{cora_sag_2018}. It simulates the formation and growth of galaxies within
each detected dark matter halo, tracking the evolution of galaxy properties
through the merger trees of haloes and subhaloes. The model can be used to
study a wide range of galaxy properties, including their luminosity functions,
mass functions, star formation histories, and morphologies.  The \sag~model
incorporates physical processes such as gas cooling, star formation and
chemical enrichment.  Star formation proceeds in both quiescent and bursty
modes; the former takes place in gaseous discs, and the latter is triggered by
disc instabilities and mergers contributing to the formation of a stellar bulge
and the growth of a central supermassive black hole. Feedback processes from
both supernovae and active galactic nuclei are also included.

One of the key strengths of the \sag~model is its ability to incorporate
environmental effects, such as ram pressure stripping (RPS) and tidal
stripping, on galaxy properties. These effects occur predominantly within
groups and clusters, which is particularly relevant to the current
investigation.  Upon becoming satellites, galaxies retain a hot gas halo that
is gradually removed by various processes, with RPS being the primary
contributor. As a result, the hot gas halo of a satellite galaxy serves as a
protective barrier against the effects of the ram pressure exerted by the
intragroup/intracluster medium on the cold gas located within the galaxy's
disc. This protective function persists provided that the ratio between the hot
gas halo and the galaxy's baryonic mass is greater than 0.1. Nonetheless, if
this ratio drops below the aforementioned threshold, the hot gas halo is
considered exhausted, which enables ram pressure to strip the cold gas disc.
The application of RPS relies on a novel analytical fitting profile, which
simulates the force of ram pressure acting upon satellite galaxies in distinct
environments (characterized by the dark matter halo mass), at different
halocentric distances and redshifts \citep{VegaMartinez+2022}.

The \sag~model accounts for orphan galaxies by monitoring satellites left over
from haloes that the underlying simulation can no longer resolve. It uses the
information provided by an orbital evolution model that encompasses both
dynamical friction and mass-loss by tidal stripping \citep{Delfino+2022}.
Although the calibration of the model, that is, the fine adjustment of the free
parameters associated with certain implemented physical processes, takes orphan
satellites into account, they are not included in the sample analyzed in this
study. 

The calibration is accomplished by utilizing the Particle Swarm Optimization
technique \citep{ruiz_sag_2015}, which yields a set of best-fitting values for
the free parameters by comparing the model results against a given set of
observables. The galaxy properties considered for calibration include the
stellar mass functions at redshifts $z=0$ and $z=2$, the distribution function
of star formation rates at $z=0.15$, the percentage of cold gas mass as a
function of stellar mass at $z=0$, and the correlation between bulge mass and
the mass of the central supermassive black hole at $z=0$.  Table 1 in
\citet{cora_sag_2018} displays the values of the free parameters defining the
model version utilized in this study, with the exception of the parameter
responsible for regulating the redshift dependence of the reheated and 
ejected cold gas by supernova feedback (parameter $\beta$; see their Eqs. 10 and 12).
This parameter has been reduced to enhance the agreement between the simulated and
observed values of the evolution of the star formation rate density and the 
fraction of quenched galaxies as a function of stellar and halo mass, which are 
predictions of the model (see, respectively, their Figs. 6 and 11).
During the calibration process, the fit to the stellar mass function at $z = 2$ 
results in $\beta = 1.99$, which favours larger supernova feedback efficiency at 
higher redshifts and, consequently, a reduction of the star formation activity at 
those redshifts; this activity is shifted to later epochs and $z=0$ galaxies have 
less time to be quenched. A better agreement between the aforementioned model 
predictions and observational data is obtained by fixing $\beta = 1.3$ while keeping 
the rest of the parameter values from the calibration process, at the expense of 
predicting a higher number density of low-mass galaxies in the stellar mass function 
at $z = 2$. This smaller value of $\beta$ is provided by the {\sc fire} hydrodynamical 
simulations of \citet{Muratov_2015}, which guided the current implementation of 
supernova feedback in \sag.


\subsection{The sample of simulated backsplash galaxies, three stages, four types}
\label{sect:BS}

Our study focuses on 34 massive, relaxed, and isolated galaxy clusters that
were selected from the \mdpl-\sag~galaxy catalogue. To identify these clusters,
we first applied selection criteria based on the halo mass and the presence of
nearby companion haloes. Specifically, we selected all haloes at redshift $z=0$
with a mass $M_{\rm 200}\ge 10^{15}\ \hmsun$, and with no companion haloes
within $5\times R_{\rm 200}$ that are more massive than $0.1\times M_{\rm
200}$. Here, $M_{\rm 200}$ refers to the mass enclosed within a region of
radius $R_{\rm 200}$ that encompasses 200 times the critical density. These
selection criteria were designed to exclude haloes that are undergoing major
mergers or interacting with massive companions, which may have an impact on the
orbits of galaxies in the vicinity of the clusters. This cluster sample is
identical to the one employed in the works of \citet{roger} and
\citet{Coenda:2022} for the development  and testing of the \textsc{roger} code
which dynamically classifies galaxies in and around clusters in the projected
phase space.  Throughout this paper we use galaxies with $M_{\star} \geq
3\times10^8\ \hmsun$ since stellar mass functions and galaxy properties cannot
be reliably replicated for galaxy masses lower than this threshold
\citep{Knebe2018}. 

We have adopted the same classification scheme as \citet{roger} to categorize
galaxies in and around the selected clusters based on their orbits.  In
particular, we define backsplash (BS) galaxies as those galaxies which at $z=0$ are
found outside $R_{200}$, having crossed this radius exactly twice in their
lifetimes, the first time on their way in, and the second on their way out of
the cluster. 

Hereafter, our paper will solely focus on the backsplash galaxies, and we will 
refer to them as BS galaxies to prevent confusion with the term
\textit{backsplash stage}, introduced below.

We follow the evolution of galaxies through three stages in their lifetimes:
\begin{itemize}
\item 
\textbf{Incoming stage:} 
the period of $2\,{\mathrm{Gyr}}$ preceding the first time a galaxy crosses 
$R_{\rm 200}$ in an inward direction.
\item
\textbf{Diving stage:} the span of time the galaxy remains within $R_{\rm 200}$.
\item
\textbf{Backsplash stage:}  
the lapse of time between the moment when the galaxy crosses $R_{\rm 200}$ for 
the second time in an outward direction, and the present epoch at $z=0$.
\end{itemize}

Following the criterion adopted by \citet{cora_sag_2018}, galaxies with a
specific star formation rate (sSFR) higher than $10^{-10.7}$~yr$^{-1}$ are
defined as star forming, meanwhile galaxies with sSFR below that threshold are
classified as passive.  We have classified the galaxies into four distinct
types based on whether they are star forming or passive at the different
stages.

\begin{figure}
\includegraphics[width=\columnwidth]{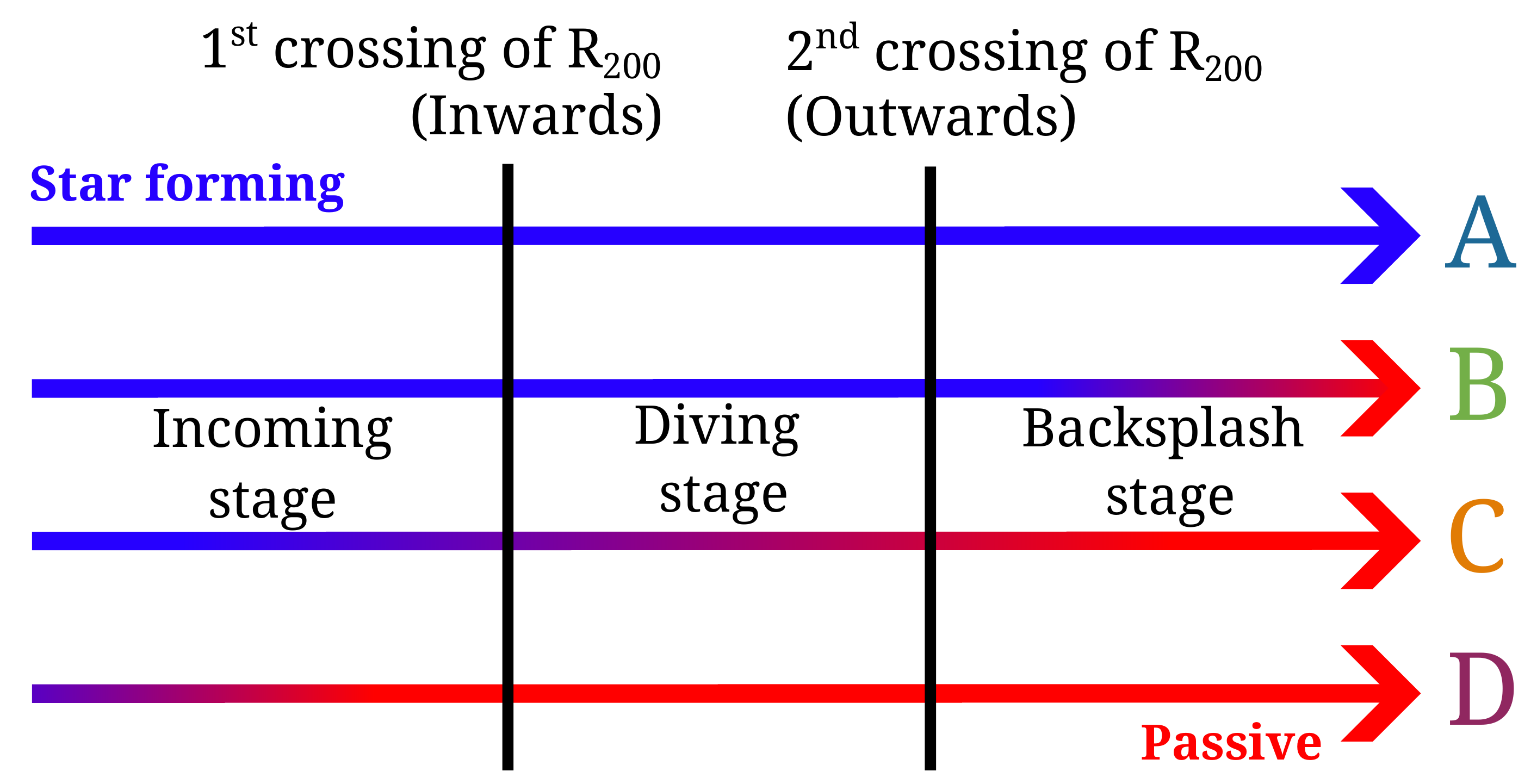}
\caption{Scheme adopted for the selection of different types of
	BS galaxies based on the
	moment when a galaxy transitions from being star forming to a passive
	state.  Blue colour symbolises a star forming galaxy, while red denotes
	a passive one.  The four types of galaxies are: galaxies that never
	become passive (A); galaxies that become passive during their
	backsplash stage (B); galaxies that become passive while they are
	inside the cluster (C); galaxies that become passive before entering
	the cluster (D).}
\label{fig:scheme}
\end{figure}

\begin{itemize}
\item \textbf{Type A:} galaxies that are star forming throughout the three 
stages; 3038 galaxies in total, 58 per cent of the sample.
\item \textbf{Type B:} galaxies that become passive during their backsplash 
stage; 1353 galaxies, 26 per cent of the sample.
\item \textbf{Type C:} galaxies that become passive during their diving 
stage; 200 galaxies, 4 per cent of the sample.
\item \textbf{Type D:} galaxies that become passive 
prior to their diving stage, i.e. during the incoming stage or even before;
622 galaxies, 12 per cent of the sample.
\end{itemize}
This classification scheme is summarised in Fig. \ref{fig:scheme}.

\begin{figure}
\includegraphics[width=\columnwidth]{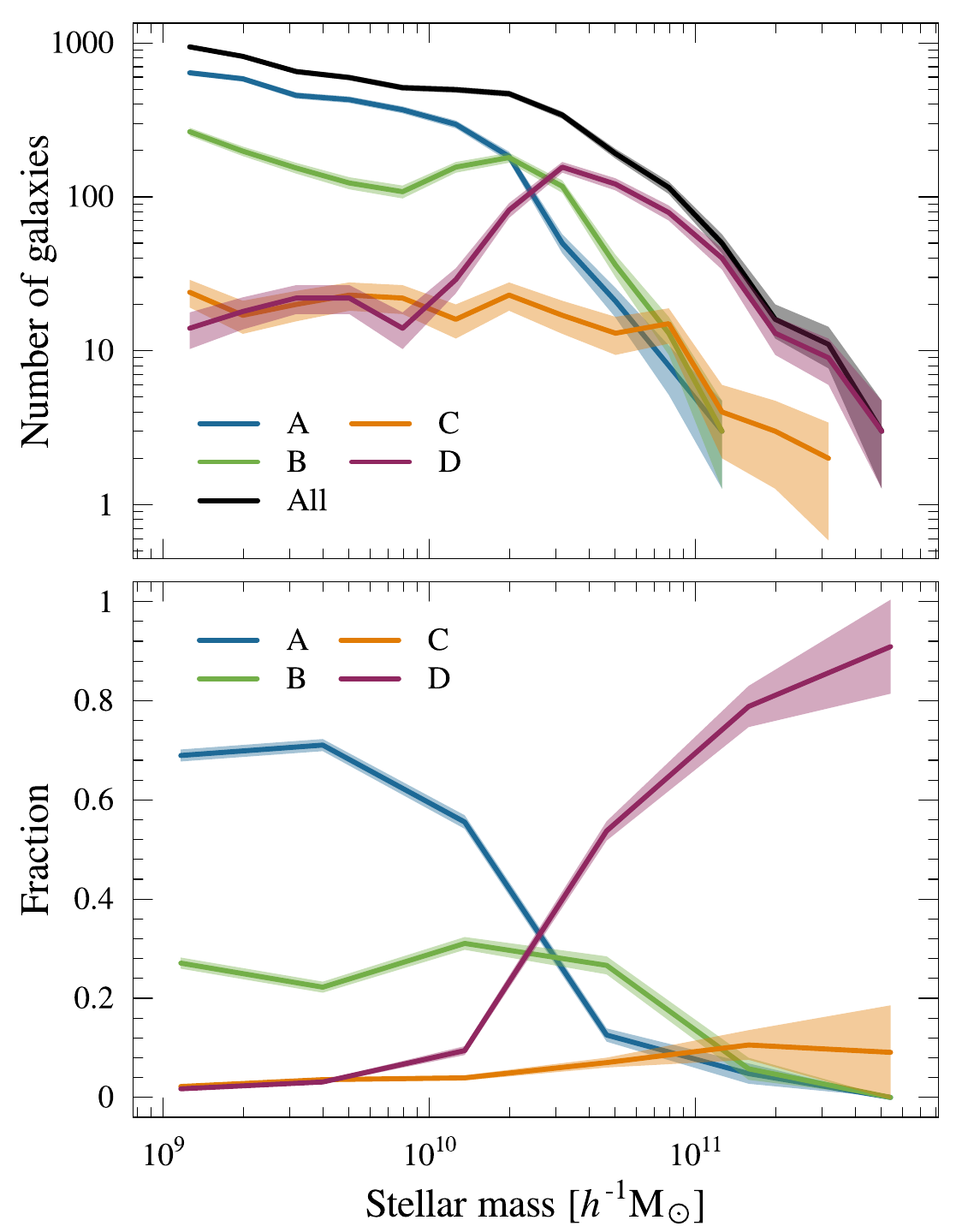}
\caption{{\it Upper panel:} Stellar mass distributions for the four types 
         (color lines) and for the complete sample (black line) of BS galaxies 
         at $z=0$. Shaded regions represent Poisson errors.
         {\it Bottom panel:} Fractions of the four galaxy types as a function of 
         their stellar mass at $z=0$. The median values are indicated by lines, 
         while bootstrap errors are represented by shaded regions.}
\label{fig:mst}
\end{figure}

In Fig. \ref{fig:mst}, we show both the stellar mass distributions 
for the four types types and for the total sample of BS galaxies at $z=0$ 
(upper panel), and the fraction of galaxies of each type as a function of 
their stellar mass at $z=0$ (bottom panel). There is no surprise in this plot. At
the low-mass end ($M_{\star}\lesssim 10^{10} \hmsun$), the majority of galaxies
are still star forming (type A).  The fractions that follow in descending order
correspond to galaxies that ceased star formation after leaving the cluster
(type B).  On the other hand, at the high-mass end ($M_{\star}\gtrsim 10^{11}
\hmsun$), most of the galaxies were passive prior to diving into the cluster
(type D). The transition between these two regimes occurs at intermediate
masses, $M_{\star} \sim 3 \times 10^{10} \hmsun$.  The fraction of galaxies
that became passive during their diving stage (type C) is very low over the
whole range of mass, accounting only for 4 per cent of the complete sample of
galaxies. This is an indication that the cluster environment can not easily
quench a galaxy while diving though it for the first time.  Nevertheless, about
a quarter of all BS galaxies ($\sim 30$ per cent of the galaxies that
entered as star forming), become passive after leaving the cluster (Type B),
which suggests that the effects of the cluster environment on the star
formation of galaxies are not immediate and may take some time to manifest.
Galaxies with stellar masses below the aforementioned transition value exhibit
a clear manifestation of this phenomenon. To gain deeper insights into the
underlying physical mechanisms responsible for the variation in the proportions
of different galaxy types and their correlation with stellar mass, we analyse
in detail  the dynamics and astrophysical properties of these objects.


\section{Dynamics}
\label{sec:dynamics}

This section delves into the main aspects of the dynamics of BS
galaxies, which comprises their entry and exit points from clusters,
velocities, duration of the diving stage, and the proximity to the cluster
centre at their pericentre.  It is essential to consider not only distances
with respect to the centers of clusters but also directions in space concerning
the main axes of the clusters. This approach is crucial for a comprehensive
understanding of the spatial orientation of objects within the cluster
environment.  We determine the main axes of each cluster in our sample by
diagonalising the cluster's shape tensor.  For this purpose, we use the
position of all galaxies whose dark matter haloes are subhaloes gravitational
bound the the main cluster halo as identified by \textsc{Rockstar} at $z=0$,
that is satellite galaxies within dark matter substructures.  The shape tensor
is defined as:
\begin{equation} 
S_{ij}  =  \sum_{k=1}^{N} X_k^i X_k^j, \label{eq:shape_tens}
\end{equation}
where the dummy index $k$ runs from 1 to the number of galaxies, $N$, and
$X^i_k$ and $X^j_k$ are the $i-$axes and $j-$axes Cartesian coordinates of the
galaxy $k$, respectively.  By solving the eigenvalue problem of the shape
tensor, we determine the direction of the orthogonal axes of symmetry of the
galaxy distribution. The main axis will be the one along which the distribution
of galaxies is most extended. The secondary and tertiary axes follow in
decreasing elongation.  We have computed the main axes of the clusters at many
different outputs of the \sag~model, and found that they are stable enough for
our purposes in the last few Gyrs. Thus, for simplicity, we consider the main
axes at $z=0$ in our computations. 

For each cluster we define a right hand rule Cartesian coordinate system where
the $X-$axis, $Y-$axis and $Z-$axis are the main, secondary and tertiary axes,
respectively.  We define clustercentric angular positions relative to these
axes taking as fundamental plane the $X-Y$ plane: the longitude $l\in[0^\circ,
360^\circ)$, and the latitude $b\in [-90^\circ,90^\circ]$.  We compute the
angular position of each galaxy in our sample relative to its parent cluster
and stack them all into an unique sample we use throughout the paper.

\begin{figure}
\includegraphics[width=\columnwidth]{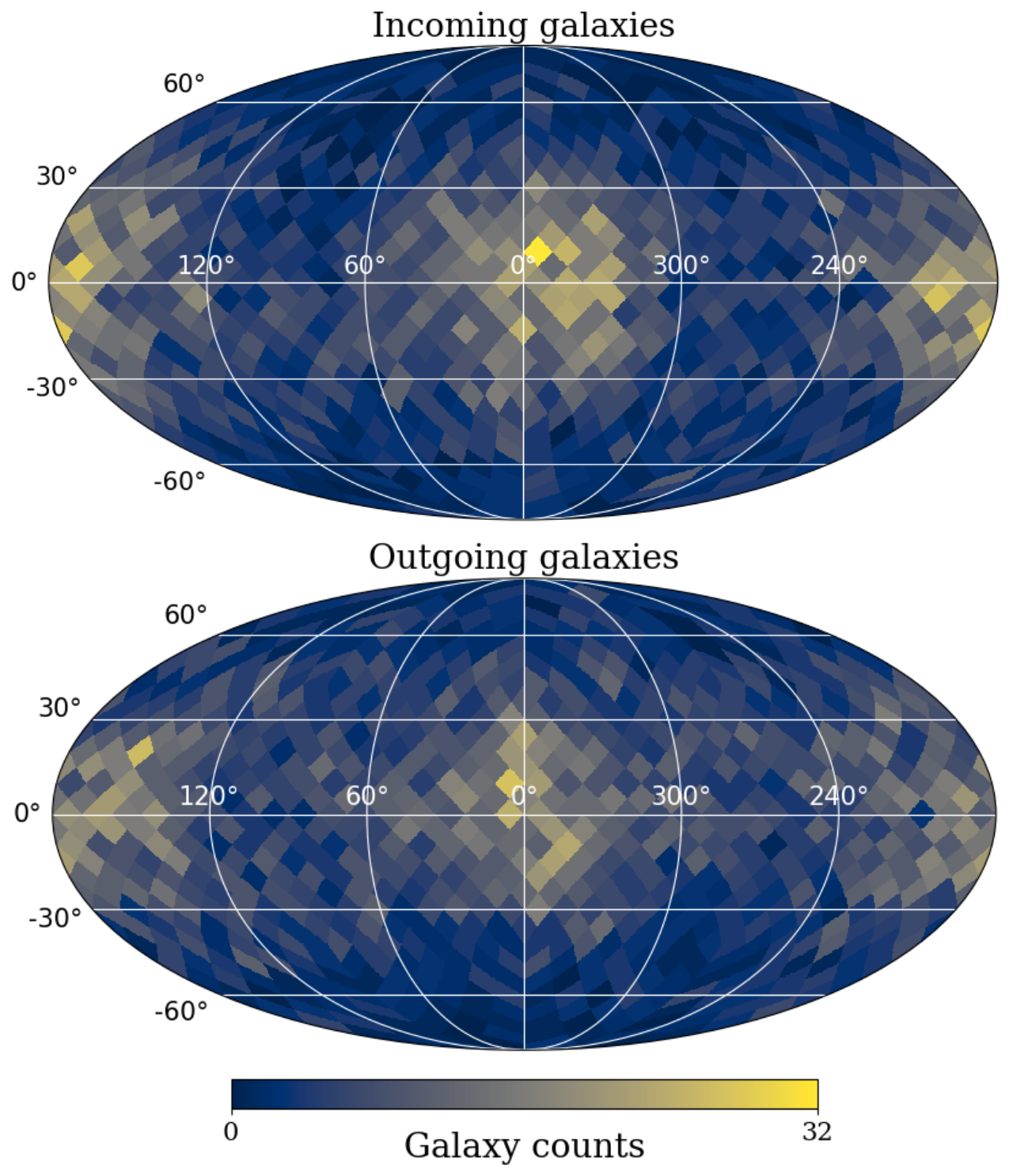}
\caption{Mollweide projection of the angular distribution of galaxies relative
	to their parent cluster's main axes. \textit{Upper panel:} the angular
	distribution of incoming galaxies at the moment they cross $R_{200}$
	of the cluster in an inward direction.  \textit{Lower panel:} the
	angular distribution of outgoing galaxies as they move away from the
	$R_{200}$ boundary.  In these plots, the coordinates of the main axis
	are $(l,b)=(0^{\circ},0^{\circ})$, and $(180^{\circ},0^{\circ})$; the
	secondary axis' coordinates are $(l,b)=(90^{\circ},0^{\circ})$, and
	$(270^{\circ},0^{\circ})$; the tertiary axis is located at latitude
	$b=-90^{\circ}$, and $90^{\circ}$. }
\label{fig:bs_ang}
\end{figure}

We show in Fig. \ref{fig:bs_ang} the stacked angular positions of galaxies when
they enter their parent cluster and when they leave it (upper and lower panels,
respectively).  In both cases, there is a noticeable accumulation of galaxies
around the primary axis. However, when galaxies enter the cluster, there
appears to be a stronger concentration that tends to diffuse primarily over the
$X-Y$ plane as they depart.  To quantify this, we compute the angular
overdensity of galaxies at the times they cross $R_{200}$,
\begin{equation}
    \Delta(\lambda,\beta)=\frac{N(\lambda,\beta)}{N_R(\lambda,\beta)}-1,
\end{equation}
by counting the number, $N(\lambda,\beta)$, of galaxies with angles
\begin{equation}
    \lambda = \left\{
	       \begin{array}{ll}
		 l,             & \mathrm{if\ } l \le 90^\circ \\
		 |180^\circ-l|, & \mathrm{if\ } 90^\circ < l < 270^\circ \\
		 360^\circ-l,   & \mathrm{if\ } l \ge 270^\circ ,
	       \end{array}
	     \right.
\end{equation}
and 
\begin{equation}
    \beta = |b|.
\end{equation}
These angles are the distance in longitude with respect to the cluster main
axes, and the distance in latitude from the cluster main plane.
$N_R(\lambda,\beta)$ is the expected number of points with angular coordinates
$(\lambda$, $\beta)$ in an homogeneous angular distribution over the sphere.
To compute this number we generate a random distribution of angular points which
covers the sphere homogeneously. This distribution contains 100 times the number of
galaxies in the sample, therefore, $N_R(\lambda,\beta)$ is the number of
these random points with  coordinates $(\lambda,\beta)$ divided by 100.
In addition, we compute the overdensity of
galaxies, $\omega(\theta)$, as a function of the angular distance to the main
axis, $\theta=\arccos(\cos(\beta)\cos(\lambda))$, which is computed in an
analogous way as $\Delta(\lambda,\beta)$.

\begin{figure*}
\includegraphics[width=0.9\textwidth]{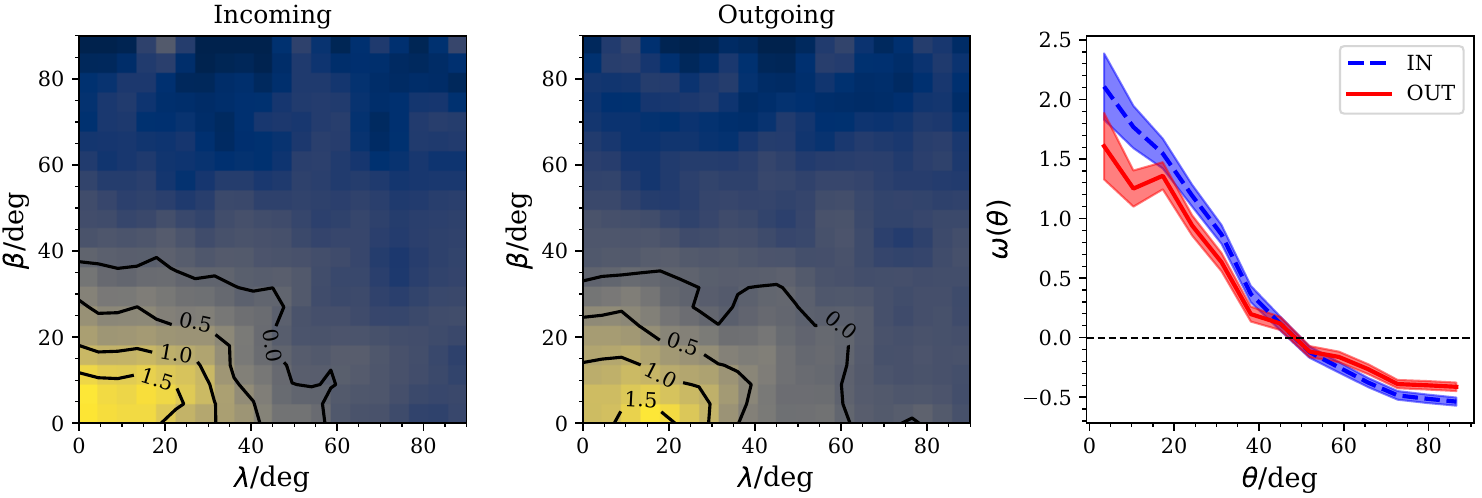}
\caption{Angular overdensity of  galaxies relative to their parent cluster's
	main axes. \textit{Left panel:}  overdensity of incoming galaxies as
	they enter $R_{200}$ as a function of the distance in longitude with
	respect to the  primary axis, $\lambda$, and the absolute value of
	latitude, $\beta$.  In this plot, the  primary axis of the cluster has
	coordinates $(\lambda,\beta)=(0^\circ, 0^\circ)$, the secondary
	$(\lambda,\beta)=(90^\circ, 0^\circ)$, and the tertiary
	$\beta=90^\circ$. Some relevant isocontours are drawn in solid line.
	\textit{Middle panel:} same as the \textit{left panel} but for outgoing
	galaxies. \textit{Right panel:} the angular overdensity of incoming
	(IN) and outgoing (OUT)  galaxies as a function of the angular
	distance, $\theta$, to the  primary cluster axis. Errors are computed
	using the bootstrap resampling technique.}
\label{fig:tpcf}
\end{figure*}
\begin{figure*} 
\includegraphics[width=0.9\textwidth]{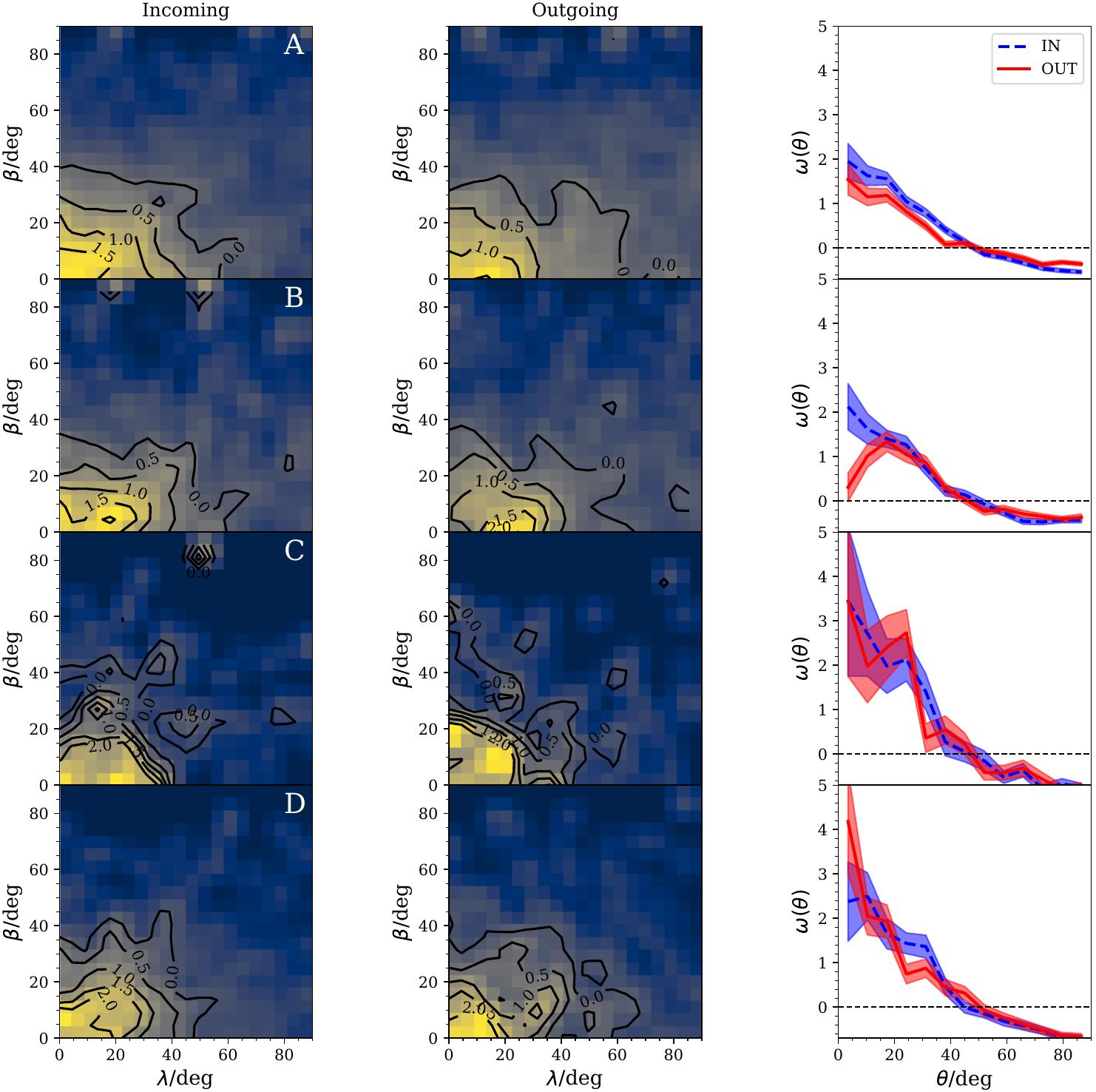}
\caption{Same as Fig.~\ref{fig:tpcf} but considering each of the four
         types of galaxies separately as quoted in \textit{left panels.}}
\label{fig:tpcf_clas}
\end{figure*}
\begin{figure*}
\includegraphics[width=1.75\columnwidth]{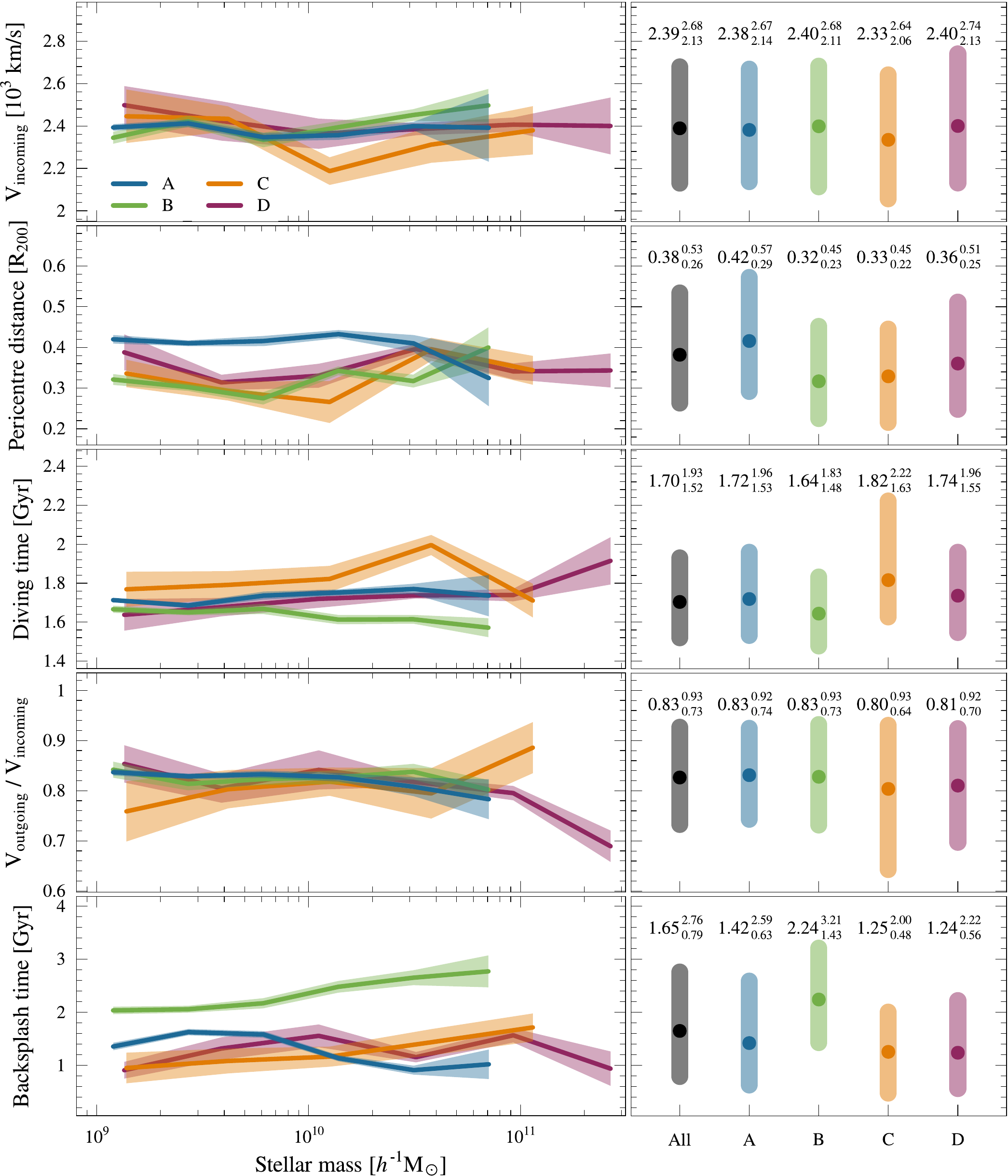}
\caption{{\it Left panels:} Dependence of several dynamical properties with
	stellar mass for the four types of galaxies. From top to bottom:
	incoming velocity modulus, normalized pericentre distance, time in the
	diving stage, ratio between outgoing and incoming velocity modulus and
	time in the backsplash stage. In all cases we show the median value and
	the errors computed with boostrap resampling. {\it Right panels}:
	Median values (central dot) and quartiles (shaded bars) for the same
	properties described in left panels. Numerical values are quoted for
	each type, including the complete sample of galaxies.}
\label{fig:prop_mst}
\end{figure*}

The results of the computation of the angular overdensity for the general
population of galaxies are shown in Fig.~\ref{fig:tpcf}.  The left panel of
this  figure shows that the majority of galaxies tend to enter the cluster
within a range of $\sim 35^\circ$ of the plane, with some entering up to $\sim
50^\circ$ in longitude; a strong concentration is observed towards the primary
axis itself. The overdensity isocontours are found to extend farther in
longitude than in latitude, with a difference of a factor $\sim 2$ for the two
highest value isoncontours shown ($\Delta=1$, and 1.5).  For outgoing galaxies
(central panel), changes in the shape of the isocontours are observed, with the
highest overdensity isocontours ($\Delta\geq 0.5$) appearing to shrink.
However, the overall overdensity enclosed by the $\Delta=0$ isocontour is found
to extend further over the main plane, with a range of up to $\lambda\sim
60^\circ$.  The majority of galaxies  tend to enter the cluster near the
primary axis and over the main plane, and to exit the cluster over the same
plane, albeit with less concentration towards the  primary axis.  The  right
panel of Fig.~\ref{fig:tpcf} reinforces this behaviour: while the overdensities
of both incoming and outgoing galaxies are seen at angular scales of $\theta
\lesssim 50^\circ$, the overdensities of incoming galaxies are systematically
greater over that range.  This deflection is in qualitative agreement with the
results by \citet{Kuchner:2022} where the authors suggest that BS entering
through a filament do not necessarily return along a filament.

We perform the same analysis for the subsamples of galaxies of different type,
as defined in the previous section. This is shown in Fig.~\ref{fig:tpcf_clas}.
Each row in this figure considers a different galaxy type, from A to D, moving
from  top to bottom respectively.  Comprising $58$ per cent of the total
sample, type A galaxies show angular overdensities that resemble those of the
entire sample, i.e., they are mildly deflected over the main plane.  Type B
galaxies constitute the most interesting case, as they also display deflections
over the main plane, but their trajectory out of the cluster systematically
deviates from the primary axis.  This is further highlighted by the
characteristics exhibited in the angular overdensity $\omega(\theta)$: outgoing
type B galaxies present a broad peak at $15^\circ\lesssim \theta \lesssim
35^\circ$, and a sharp decline towards $\theta=0^\circ$.  This strong
deflection is not seen for the other three types.  Finally, the less numerous
type C and D galaxies seem to get in and out the cluster much more closer to
the primary axis, as if they were accreted into the cluster by inner regions of
filaments and expelled out the cluster in a similar way
\citep{gonzalez_filament_2016,salerno_quenching_1,salerno_quenching_2,morinaga_filament_2020}.
This is consistent with the definition of these two types, they are either
already quenched when  they enter the cluster, or are quenched in their
incursion through it.

Other dynamical parametres are shown in Fig. \ref{fig:prop_mst}, namely:
incoming velocity, the distance of the pericentre, the timespan of the diving
stage, the ratio between the incoming and the outgoing velocities, and the
timespan of the backsplash stage. All these parameters are shown as median
values per type, as a function of stellar mass  (left panels), and as a single
value stacking all galaxies together regardless of their mass (right panel). It
is evident from the left panels that there are no notable patterns between
these parameters and stellar mass and, also, it is fair to notice the lack of
some types of galaxies for stellar masses higher than $\sim 7\times 10^{10}
\hmsun$, where a complete comparison can not be done.

Regarding the incoming velocity (top panels), it seems that Type C galaxies
have a tendency to enter clusters at lower velocities compared to the other
types, whereas no discernible difference is observed among types A, B, and D.
Instead, in the case of pericentre distance (shown in the second panel from the
top), type A galaxies stand out, as they exhibit systematically larger
distances from the cluster centre at their closest point to it than the other
galaxy types.  On the other hand, there are no significant differences among
the other types.  When analysing how long galaxies are inside $R_{200}$
(middles panels), we find that types A and D cannot be distinguished, but type
B spends the least amount of time, while on the opposite end, type C remains
for the longest duration.  There is no difference among the four types with
regards to the ratio between the outgoing and incoming velocities (second panel
from the bottom).  All types get out the cluster at speeds $\sim 20$ per cent
lower than they had when they entered the cluster.  Finally, type B galaxies
are the ones that have spent the longest time in the backsplash stage, while no
differences are seen among the other types (bottom panel).

We have checked that 27 per cent of the complete sample of BS galaxies are
falling back to to the cluster at $z=0$, and this fraction does not vary
significantly with the galaxy type. The only exception are type B galaxies for
which this percentage increases to 38, in consistency with the fact that they
spend typically longer times in the backsplash stage.


\section{The physical evolution of backsplash galaxies}
\label{sec:astrophysics}

In this section we study how physical properties of \mdpl -\sag\  BS
galaxies evolve in the three stages. We focus on some quantities provided by
the \sag\ model: stellar mass, dark matter halo mass, cold gas mass, hot gas
mass, and the specific star formation rate. Towards the end of this section, we
examine morphological changes experienced by these galaxies by using the ratio
of bulge mass to total stellar mass as a proxy of morphology.


\subsection{Astrophysical quantities}

\begin{figure*}
\centering
\includegraphics[width=\textwidth]{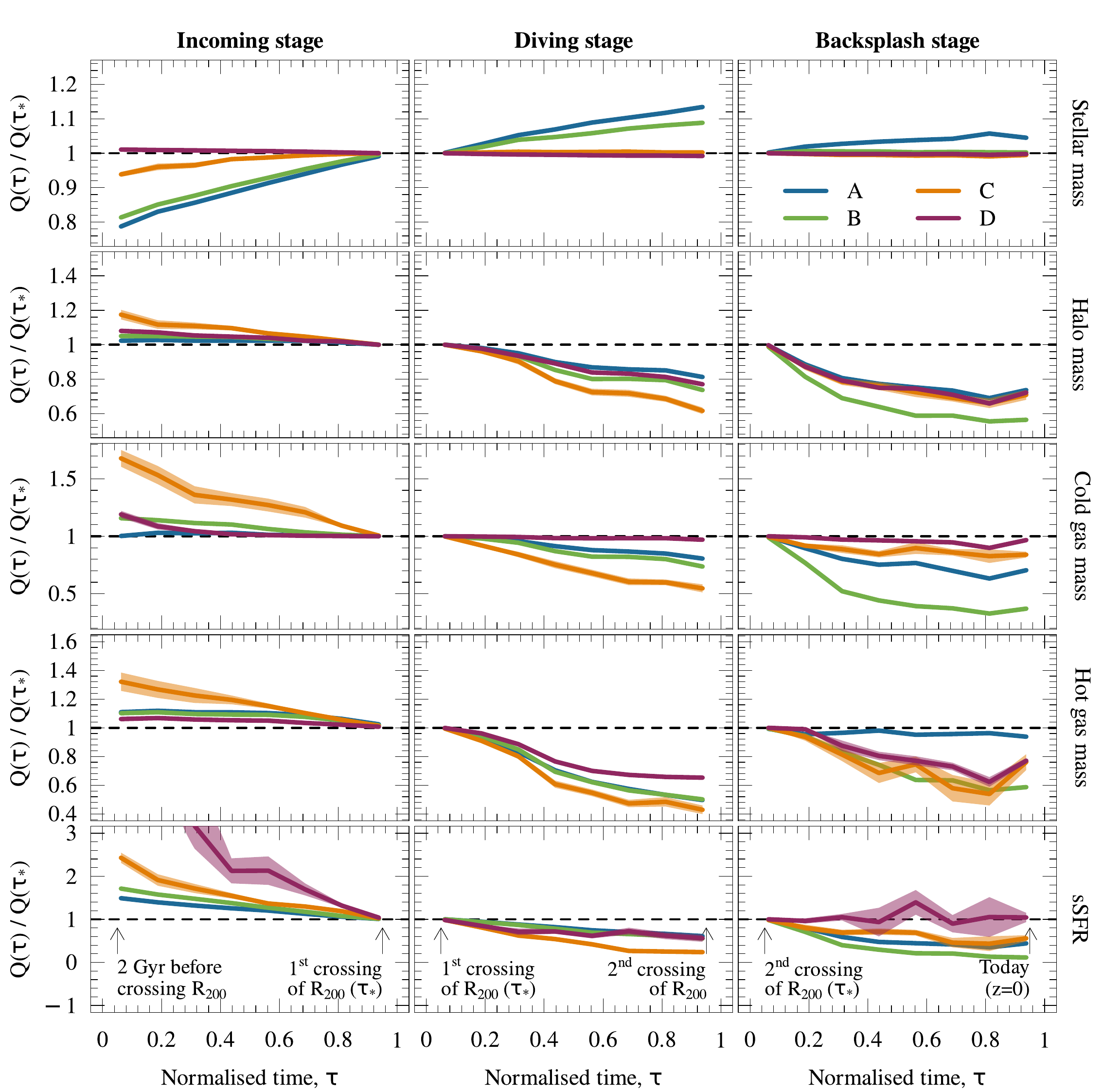}
\caption{Evolution of normalised astrophysical properties for the four types of
	galaxies in the three stages. Evolution is parameterized with $\tau$, a
	normalized time at two particular moments in each stage, as indicated
	with arrows at the bottom panels. The astrophysical properties are
	normalised at their values at $\tau_\ast$, which is the first crossing of
	$R_{200}$ for the incoming and diving stages, and the second crossing
	of $R_{200}$ for the backsplash stage. Properties $Q$ are, from top to
	bottom, the stellar mass, the dark matter halo mass, the mass in cold
	gas, the mass in hot gas and the specific star formation rate. In all
	cases, we show the median of the stacked galaxy population and
	errorbars computed via boostrap respampling.}
\label{fig:evolution}
\end{figure*}
\begin{figure*}
\includegraphics[width=0.9\textwidth]{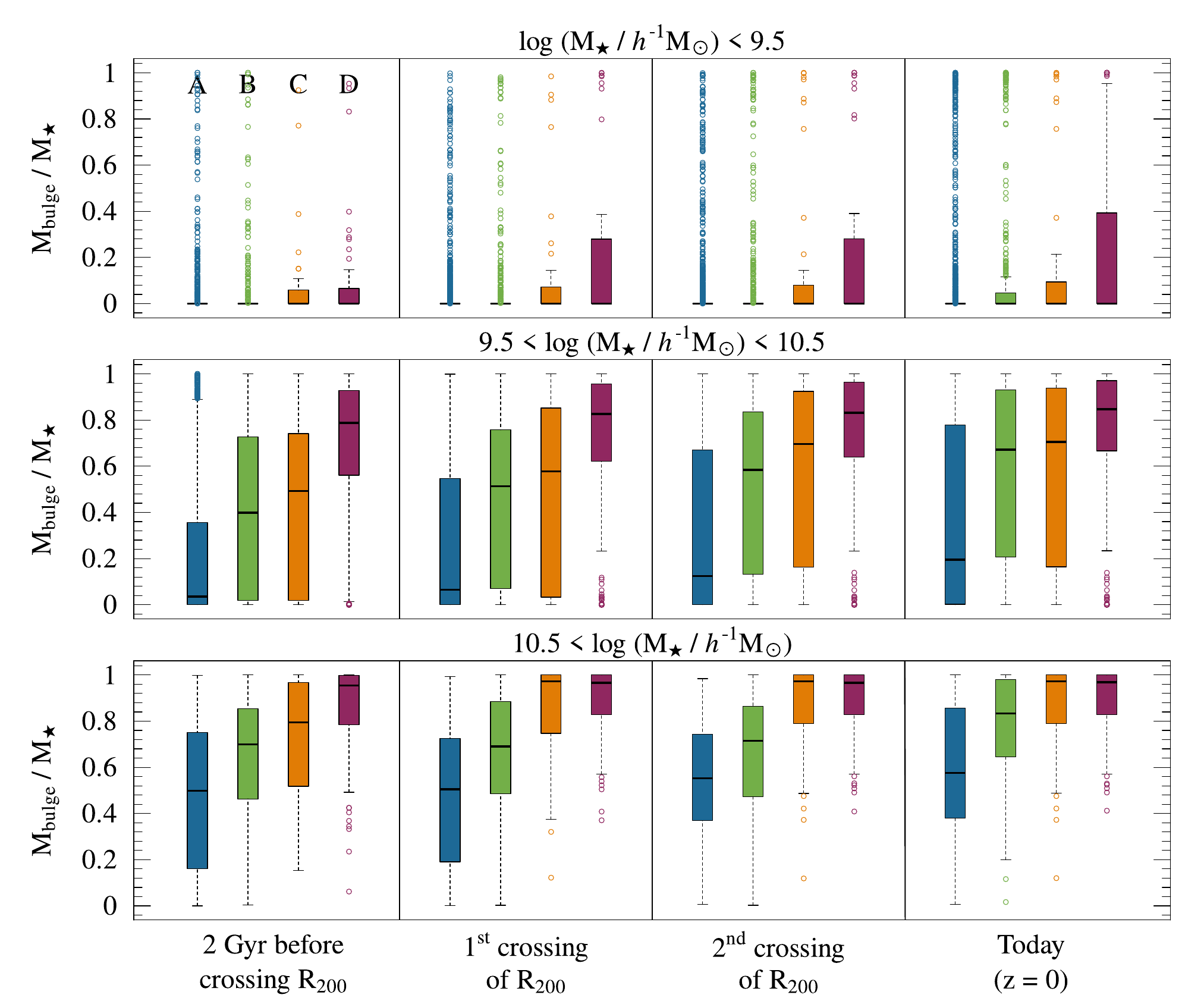}
\caption{Boxplots for morphological distributions for the four galaxy types,
	where morphology is defined as the ratio between the bulge mass and the
	total stellar mass, $M_{\rm bulge} / M_{\star}$. Distributions are
	shown for three stellar mass cuts, from low-mass galaxies (top) to
	high-mass galaxies (bottom), and at four different moments in the
	galaxy lifetime, as indicated in the $x$-axis labels. In each boxplot,
	the color bar represent the interquartile range $IQR = Q_3 - Q_1$,
	where $Q_1$ and $Q_3$ are the first and third quartiles of the
	distribution, respectively, and the median value is shown with the
	short horizontal black line. The dashed vertical lines show the range
	covered by the minimum and maximum values, without considering the
	outliers of the distribution, which are defined as those at a distance
	greater than $1.5 \times IQR$ of $Q_1$ or $Q_3$ and symbolized with the
	coloured open circles.}
\label{fig:morphology}
\end{figure*}

The evolution of astrophysical properties of BS galaxies is presented in Fig.
\ref{fig:evolution}.  As temporal variable we use $\tau$, defined as the time a
galaxy remains in a particular stage conveniently normalised. For the incoming
stage, we considered as $\tau=0$ to 2 Gyr before the galaxy crosses $R_{200}$
inwards the cluster, and $\tau=1$ as the moment when the galaxy crosses
$R_{200}$ for the first time.  For the diving stage, $\tau=0$ is set at the
moment the galaxy crosses $R_{200}$ inwards the cluster and $\tau=1$ when the
galaxy crosses $R_{200}$ outwards.  Finally, in the backsplash stage, we define
as $\tau=0$ the time the galaxy crosses $R_{200}$ for the second time and
$\tau=1$ corresponds to $z=0$. With this normalised time, we stacked the
evolution of astrophysical properties normalised to their values at a
particular moment $\tau_{\ast}$: this $\tau_\ast$ is the first crossing of
$R_{200}$ for incoming and diving stages, and corresponds to the second
crossing of $R_{200}$ for the backsplash stage. All these time definitions are
shown at the bottom panels of the figure.  From top to bottom in Fig.
\ref{fig:evolution}, the properties shown are: stellar mass, dark matter halo
mass, cold gas mass, hot gas mass, and specific star formation rate.  In all
cases, we show the median values of the evolution of the stacked galaxy
population with errorbars (shaded regions) computed with a boostrap resampling.

We begin our analysis by studying the total stellar mass of galaxies. Type A
galaxies exhibit the most substantial increase in their stellar masses across
the three stages: approximately $20$ per cent during the incoming stage, $10$
per cent  during the diving stage, and $5$ per cent during the backsplash
stage. Type B galaxies show a similar trend during the first two stages, but
their median growth in stellar mass halts during the backsplash stage.  This
outcome is anticipated since, by definition, these galaxies become passive
during this stage.  In the case of types C and D galaxies, only type C galaxies
exhibit a marginal increase of $\sim 5$ per cent in their masses during the
incoming stage.  These behaviours are consistent with our classification:
galaxies that become passive at a later stage exhibit a correspondingly delayed
decrease in the median stellar mass growth.

The evolution of the mass of the dark matter halo where the galaxy resides is
completely determined by the dynamical interactions of the haloes across their
merger trees histories. As can be seen in the second row of panels from
Fig.~\ref{fig:evolution}, for all galaxy types, the host haloes decrease in
mass across the three stages. In the incoming stage, haloes of type C galaxies
are the most affected, loosing up to $\sim 20$ per cent of their mass.
Similarly, the diving stage causes a more significant impact on the haloes of
type C galaxies ($\sim 40$ per cent),  whereas the haloes of the remaining
types experience losses of around 20 per cent.  The fact that type C galaxies
become passive during the diving stage may be correlated with the significant
mass disruption experienced by their host haloes during this stage.  This tidal
stripping of their haloes exposes them to a stronger effect of the ram pressure
upon their hot gas (see \citealt{VegaMartinez+2022} for details of the
implementation of these procesess in \sag). Furthermore, these galaxies endured
heavy pre-processing in the incoming stage.  During the backsplash stage,
haloes of type B galaxies are those that experience the greatest impact on
their masses, with a decrease of over $\sim 40$ per cent. Nevertheless, haloes
of the other galaxy types also show a reduction of $\sim 30$ per cent in their
masses at $z=0$. Once again, we notice a correlation between the haloes that
suffer the most significant mass loss and the fact that the galaxies inhabiting
them (type B) become passive during this stage.  This finding is consistent
with their null stellar mass growth. It is worth noting that, during the
backsplash stage, all galaxies except for type A have ceased to grow their
stellar mass, and the dark matter haloes continue to shrink across all
classifications.

In the third and fourth rows of panels of Fig.~\ref{fig:evolution}, we show the
evolution of the cold and hot gas mass, respectively. Type A galaxies do not
undergo significant changes in their cold and hot gas masses during the
incoming stage, but suffer a reduction of $\sim 20$ per cent of the cold gas
and $\sim 50$ per cent of the hot gas when they are located inside the
$R_{200}$ radius of the cluster. Once they are out, in the backsplash stage,
they continue losing cold gas ($\sim 30$ per cent) but the amount of hot gas
remains stable.  This may be explained by the negligible effect of the RP on
these galaxies due to their greater distance from the cluster.  Type B galaxies
lose more than $50$ per cent of their cold gas during the backsplash stage,
period where these galaxies become passive. They also get their hot gas
drastically removed in the diving ($\sim 50$ per cent) and backsplash ($\sim
40$ per cent) stages.  Notably, the difference between type B and type A
galaxies is that in the backsplash stage, type B galaxies experience a
considerable removal of hot gas, while type A galaxies retain their hot gas.
This difference explains why type B galaxies become passive during the
backsplash stage.  In the incoming stage, type C galaxies suffer a large
reduction of $\sim 70$ per cent in their cold gas mass, which continues during
the diving stage, reaching another reduction of $\sim 50$ per cent at the time
the galaxy leaves the cluster. The same occurs with hot gas, suffering
reductions of $\sim 30$, $60$ and $40$ per cent during the incoming, diving and
backsplash stages, respectively.  Type D galaxies exhibit a notable change in
their cold gas content only during the initial phase of the incoming stage
which is in line with their quenched state upon entering the cluster. At this
point, the cold gas is neither consumed by star formation nor removed by RPS,
as these galaxies still possess a hot gas shield that is gradually stripped
away, reaching a loss of $\sim 35$ per cent in the diving and backsplash
stages.

Finally, in the last row we show the sSFR. While all galaxy types exhibit a
decrease in their sSFR throughout all stages, it is evident that type D
galaxies, which are already passive upon infalling into the cluster, are highly
affected during the incoming stage. These galaxies experience a further
decrease in their sSFR during the diving stage and maintain a stable sSFR
during the backsplash stage.  Type C galaxies also experience a significant
drop in their sSFR during the incoming stage, and they are the most affected
among all types in the diving stage, losing nearly all their ability to form
stars.  During the incoming and diving stages, types A and B exhibit similar
trends by reducing their sSFR by $\sim 50$. However, in the backsplash stage,
type B galaxies become passive and experience the most significant decrease in
sSFR.


\subsection{Morphology}

The overall analysis of the sSFR fails to distinguish between the two modes of
star formation: quiescent and bursty. In the context of \sag, generally
considered in galaxy formation models, quiescent star formation contributes to
the creation of the stellar disc, while bursty star formation, triggered by
mergers and disc instabilities, leads to the formation of the stellar bulge.
The proportion of the galaxy's total stellar mass that the bulge represents is
an indicator of the galaxy's morphology.

In this section we analyse how the morphology of galaxies evolves across the
three stages. To do this, we define as a proxy of the morphology the ratio of
the bulge mass to the total stellar mass, $M_{\rm bulge} / M_\star$. This ratio
ranges between 0 for irregular galaxies (no bulge mass) to 1 for elliptical
galaxies (no disc mass), with spiral galaxies between those two extreme values. 

In Fig. \ref{fig:morphology} we present boxplots for morphology distributions
at four fixed times, from left to right: 2 Gyr before the galaxy crosses
$R_{200}$ for the first time, the moment when the galaxy crosses $R_{200}$
inwards, the moment when the galaxy crosses $R_{200}$ outwards and the present
time ($z=0$). Also, we divide the samples of galaxy types into three stellar
mass ranges, from top to bottom: $\log (M_\star/\hmsun) < 9.5$ (low mass), $9.5
< \log (M_\star/\hmsun) < 10.5$ (intermediate mass) and $10.5 < \log
(M_\star/\hmsun)$ (high mass).

The first feature easily noticeable is that, at early times, low-mass galaxies
(top panels) mostly have $M_{\rm bulge} / M_\star \sim 0$, which does not
change significantly across the galaxy lifetime.  In particular, types A and B
populations are highly dominated by bulgeless galaxies, which represent $90$,
$84$, $81$ and $79$ per cent of the sample for type A, and $85$, $79$, $75$ and
$71$ per cent of the sample for type B, at each of the moments considered along
the galaxy lifetime. Only type D galaxies show a mild evolution in their
morphology, specially between the  $2\,\mathrm{Gyr}$ before of entering the
cluster and the moment galaxies crosses $R_{200}$ for the first time, that is,
during the incoming stage. For intermediate and high stellar masses (middle and
bottom panels, respectively), we can appreciate a smooth morphological change
across cosmic time for all galaxy types.

These results show that the morphological evolution of galaxies is a smooth and
continuous process during the stages defined in this work. We do not find
evidence that a particular moment in the lifetime of BS galaxies can induce a
major morphological transformation, in agreement with
\citet{martinez_rogerIII_2023}, where the authors find evidence that quenching
occurs faster than morphological transformation for galaxies around massive
X-ray clusters as classified with \textsc{roger} \citep{roger}.  We only find a
clear dependence of morphology with the stellar mass cuts analysed, dependence
that is also manifested through the relationship between the galaxy type and
its stellar mass (see Fig. \ref{fig:mst}).  At this point, it is worth 
recalling that larger stellar masses correlate with larger bulge masses, and this
implies the existence of larger black holes in those galaxies
\citep{mcconnell_bh_2013,kormendy_bh_2013,schutte_bh_2019}. These massive black
holes bring with an associated major feedback from active galactic nuclei that accelerates the
quenching processes, specially for types C and D.    


\section{Conclusions}
\label{sec:conclusions}

In this paper we have studied the population of BS galaxies around a
sample of massive, isolated clusters at $z=0$ in the \mdpl-\sag\ catalogue of
simulated galaxies.  It is important to recall that our definition of
BS galaxy involves a single passage within $R_{200}$, thus these
galaxies experience the environmental action of the cluster only in this single
passage. The main focus of this paper is to understand what happens to a galaxy
that passes only once through a cluster, and not the cumulative effect of
several orbits inside the cluster.  We classified the BS galaxies into
four types based on their star-forming or passive nature at three different
stages: incoming stage, diving stage, and backsplash stage. We analysed various
dynamical and astrophysical properties of these galaxies.

For galaxies that are actively forming stars at all stages (type A), we find
they typically have stellar masses below $3 \times 10^{10} \hmsun$ at $z = 0$.
Due to their orbits avoiding the innermost regions of the cluster, their
incursion into the cluster has minimal impact on them. Consequently, they do
not experience significant losses in their dark matter or gas contents,
allowing them to continue forming stars. 
    
We find that BS galaxies that become passive during the backsplash stage (type
B) also have masses typically below $3\times 10^{10} \hmsun$ at $z=0$,
similarly to type A galaxies. They constitute the vast majority of passive BS
galaxies in that mass range.  Their orbits inside the cluster are characterised
by the smallest pericentric distances of all four types, and, on average they
experience a strong deflection away from the primary axes in their way out.
This close encounter with the centre of the cluster it is the single most
important factor in the subsequent evolution of these galaxies.  The action of
the cluster leaves its mark in the strong loss of these galaxies' dark matter
and gas contents, and the subsequent suppression of their star formation in the
backsplash stage. Had these galaxies not come that close to the centre, they
would probably have been classified as Type A.  This galaxy population can be
associated to the low mass galaxies quenched after their first passage through
the pericenter in massive clusters described by \citet{wright_orbits_2022}
using the \textsc{eagle} simulations \citep{eagle_1,eagle_2}.

Regarding those galaxies that become passive during the diving stage (type C),
they are the least abundant among all BS. They entered the cluster in close
proximity to the primary axes and underwent significant pre-processing. Among
the galaxies that enter the cluster while still forming stars, these galaxies
are the ones that remain for the longest duration and endure the greatest
reductions in their dark matter and gas contents in the incoming stage.  The
cluster then provides the final blow to these galaxies' star formation.

Finally, galaxies that are passive throughout the three stages (type D) are
typically high mass galaxies that were quenched before entering the cluster.
Typically, they are accreted close to the primary axes, and get out the cluster
along the same axes. There is nothing particularly noteworthy about these
galaxies after they dive into the cluster.

Our results are in agreement with those presented by \citet{Hough:2023}, where
the authors show that $65$ per cent of present day star-forming BS galaxies did
not experience  pre-processing effects before their passage to the pericentre.
On the other hand, $70$ per cent of $z=0$ quenched galaxies suffer larger
removals of hot gas during their pre-processing instances (incoming stage) and
inside the cluster (diving stage). These quenched BS galaxies can be clearly
related with our D and C types, respectively. It is important to mention that
the authors define as BS a galaxy that is outside $R_{200}$ at $z=0$ but have
been inside the cluster at least once in the past, which is slightly different
than our definition, where we allow just one passage to the pericentre. This
difference implies some of their BS galaxies should be more quenched than ours
as they have suffered the action of the cluster in more than one passage.

Regarding morphology, a consistent pattern is observed across the four types: a
single passage through the cluster does not generally induce significant
morphological transformations, as suggested by \citet{martinez_rogerIII_2023}.
It should be kept in mind that we are using a proxy for morphology and not a
precise characterisation of it. 

The main result of this paper is that the environmental effects of the cluster
is able to affect galaxies after a single passage provided they are either:
low-intermediate mass ($<3\times 10^{10}\hmsun$) galaxies whose orbits take
them very close to the centre, or galaxies that entered the cluster heavily
pre-processed.  The former stop forming stars after exiting the cluster, while
the latter do so in situ. \\


\section*{Acknowledgements}

We kindly thank the anonymous Referee for comments and suggestions that improved 
the original manuscript. 
This paper has been partially supported with grants from Consejo Nacional de
Investigaciones Cient\'ificas y T\'ecnicas (PIPs 11220130100365CO,
11220210100064CO and 11220200102832CO) Argentina, the Agencia Nacional de
Promoci\'on Cient\'ifica y Tecnol\'ogica (PICTs 2020-3690 and 2021-I-A-00700),
Argentina, and Secretar\'ia de Ciencia y Tecnolog\'ia, Universidad Nacional de
C\'ordoba, Argentina. ANR thanks the Scaloneta for winning the 2022 FIFA World
Cup.  MdlR acknowledges financial support from the Comunidad Aut\'onoma de
Madrid through the grant SI2/PBG/2020-00005.
The \textsc{CosmoSim} database used in this paper is a service by the
Leibniz-Institute for Astrophysics Potsdam (AIP). The \textsc{MultiDark}
database was developed in cooperation with the Spanish MultiDark Consolider
Project CSD2009-00064.  The authors gratefully acknowledge the Gauss Centre for
Supercomputing e.V. (www.gauss-centre.eu) and the Partnership for Advanced
Supercomputing in Europe (PRACE, www.prace-ri.eu) for funding the
\textsc{MultiDark} simulation project by providing computing time on the GCS
Supercomputer SuperMUC at Leibniz Supercomputing Centre (LRZ, www.lrz.de). 
All analyses in this paper have been done using the {\sc fortran} language
(\url{https://fortran-lang.org}), and figures have been developed using {\sc R}
\citep{r_core}, {\sc Matplotlib} \citep{matplotlib}, {\sc HEALPix / healpy}
\citep{healpix_2005,healpy_2019} and {\sc Inkscape}
(\url{https://inkscape.org}).


\section*{Data availability}

The raw data of the semi-analytic model of galaxy formation \sag~will be shared
on reasonable request to the corresponding author.



\bibliographystyle{mnras}
\bibliography{references} 


\bsp	
\label{lastpage}

\end{document}